\documentclass[useAMS,usenatbib,usegraphicx]{mn2e}
\bibpunct[, ]{}{}{;}{a}{,}{,}
\newcommand{\cs}{c_{\rm s}}

\newcommand{\rc}{r_{\rm c}}

\newcommand{\bl}[1]{\mbox{\boldmath$ #1 $}}

\begin{document}

\title[Gravitational collapse of prestellar cores]
{The effect of non-isothermality on the gravitational collapse
of spherical clouds and the evolution of protostellar accretion}
\author[E. I. Vorobyov and Shantanu Basu]{E. I. Vorobyov$^{1,2}$\thanks{E-mail:
vorobyov@astro.uwo.ca (EIV); basu@astro.uwo.ca (SB)} and Shantanu
Basu$^{1}$ \\
$^{1}$Department of Physics and Astronomy, University of Western Ontario, 
London, Ontario, N6A 3K7, Canada \\
$^{2}$Institute of Physics, Stachki 194, Rostov-on-Don, Russia}

\date{}

\maketitle

\label{firstpage}

\begin{abstract}
We investigate the role of non-isothermality in gravitational collapse
and protostellar accretion by explicitly including the effects of 
molecular radiative cooling, gas-dust energy transfer, and cosmic ray
heating in models of spherical hydrodynamic collapse.
Isothermal models have previously shown an initial decline in the 
mass accretion rate $\dot{M}$ during the accretion phase of protostellar
evolution, due to a gradient of infall speed that develops in 
the prestellar phase. Our results show that: (1) in the idealized
limit of optically thin cooling, a positive temperature gradient is 
present in the prestellar phase which effectively cancels out the effect
of the velocity gradient, producing a near constant (weakly increasing
with time) $\dot{M}$ in the early accretion phase; (2) in the more realistic
case including cooling saturation at higher densities, 
$\dot{M}$ may initially be either weakly increasing or weakly decreasing with time, 
for low dust temperature ($T_{\rm d} \sim 6$ K) and high dust
temperature ($T_{\rm d} \sim 10$ K) cases, respectively. 
Hence, our results show that the initial decline in 
$\dot{M}$ seen in isothermal models is definitely not enhanced by 
non-isothermal effects, and is often suppressed by them.
In all our models, $\dot{M}$ does eventually decline rapidly due to
the finite mass condition on our cores and a resulting inward
propagating rarefaction wave. Thus, any explanation for a rapid
decline of $\dot{M}$ in the accretion phase 
likely needs to appeal to 
the global molecular cloud structure and possible envelope support, which
results in a finite mass reservoir for cores. 
\end{abstract}


\section{Introduction}
There is good reason to believe that the central regions of gravitationally
bound prestellar cores are somewhat cooler than their outer parts. 
This is implied by the recent far-infrared observations of starless cores
(Ward-Thompson, Andr\'e, \& Kirk \cite{WardThompson}). 
Furthermore, dust radiative transfer calculations by Zucconi, 
Walmsley, \& Galli (\cite{Zucconi}) predict that there should be a positive radial dust temperature
gradient, with values ranging from $T_{\rm d} \sim 5-7$~K at the core center to $T_{\rm d}\sim15$~K
at the core edge. Since the temperature of the gas $T_{\rm g}$ is 
coupled to that of the dust for gas densities $n>10^4$~cm$^{-3}$ 
(see e.g. Galli, Walmsley, \& Gon\c{c}alves \cite{Galli}), one may expect that the radial
distribution of the gas temperature 
in dense prestellar cores is also characterized by a positive temperature gradient.
This implies that an effective polytropic index $\gamma$ of the gas may be
less than unity, certainly during the prestellar phase.

A positive temperature gradient in the prestellar phase, and continued 
non-isothermality of a protostellar envelope during the accretion phase 
may have important consequences for protostellar evolution. 
Consider the distinct differences in self-similar solutions of spherical protostellar accretion.
Although the mass accretion rate $\dot{M}$ onto a protostar is time-independent
in isothermal similarity solutions (Shu \cite{Shu}; Hunter \cite{Hunter2}),
the similarity solutions for gravitational collapse of polytropic spheres
(Yahil \cite{Yahil}; Suto \& Silk \cite{Suto}) show that $\dot{M}$
should increase with time if $\gamma < 1$ and
decrease with time if $\gamma > 1$. The numerical modeling of the collapse
of polytropic spheres by Ogino, Tomisaka, \& Nakamura (\cite{Ogino}) also shows
this tendency. 
If the cooling in an envelope due to direct emission of photons or frequent gas-dust 
collisions is efficient enough to reduce the effective value of $\gamma$ below unity, 
then the mass accretion rate $\dot{M}$ on to the protostar is expected to increase 
with time.
However, in this case, it is difficult to explain 
the observations of Bontemps et al. (\cite{Bontemps}), who
have suggested that the mass accretion rate of Class~I objects (i.e.
protostars in the late accretion stage)
falls off by an order of magnitude compared to that of Class~0 objects (i.e. protostars 
in the early accretion phase).
One may then need to appeal to other effects that limit the mass infall, such as
a finite mass reservoir (see Vorobyov \& Basu \cite{VB}, hereafter paper I).
Conversely, the thermal balance of a protostellar envelope may be dominated
by the compressional heating of matter rather than by
radiative cooling of gas or dust-gas energy transfer. This could
raise the effective value of $\gamma$ above unity and perhaps lead to a declining
$\dot{M}$. In this case, at least some of the inferred decline in $\dot{M}$ during
the accretion phase may be attributed to non-isothermal effects.

Our motivation for this paper is to model the details of radiative cooling,
cosmic-ray heating, and gas-dust energy transfer
in a model of hydrodynamic collapse, in order to settle the 
above issues. In paper I, we studied spherical isothermal collapse and found 
that although an initially declining $\dot{M}$ is present, due to the
gradient of infall speed in the pre-stellar phase, the observational 
tracks of envelope mass $M_{\rm env}$ versus
bolometric luminosity $L_{\rm bol}$ could only be explained by 
the effect of an inwardly propagating rarefaction wave due to a finite
mass reservoir. In fact, the class~I phase of protostellar evolution 
was identified with the period of rapidly declining $\dot{M}$ occuring
after the rarefaction wave reaches the protostar.
Here, we investigate whether non-isothermality can change this 
conclusion in any way.
Specifically, can non-isothermality explain any part (or even all) 
of the inferred drop in $\dot{M}$ during the accretion phase
(Bontemps et al. \cite{Bontemps})?

The paper is organized as follows. 
The numerical code used to model the gravitational collapse, as well as
the initial and boundary conditions, are briefly discussed
in \S~\ref{model}. The temporal evolution
of the mass accretion rate for an assumption of optically thin gas 
is studied in \S~\ref{thin}. The influence of the
radiative cooling saturation and the gas-dust energy transfer (Goldsmith
\cite{Gold}) on the mass accretion rate are investigated in \S~\ref{thick}.
Our main results are summarized in \S~\ref{Sum}.

\section{Model assumptions}
\label{model}
We consider the gravitational collapse of spherical
non-isothermal clouds composed of molecular hydrogen with a $10\%$ admixture
of atomic helium. 
The evolution is calculated by solving the hydrodynamic equations in
spherical coordinates:
\begin{eqnarray}
\frac{\partial \rho}{\partial t} + \frac{1}{r^2} \frac{\partial}{\partial r} \left(r^2 \rho v_r\right)  & = & 0,\\
\frac{\partial}{\partial t} \left( \rho v_r \right) +
\frac{1}{r^2} \frac{\partial}{\partial r} \left(r^2 \rho v_r v_r\right) & = &
- \frac{\partial p}{\partial r} - \rho \frac{G\, M}{r^2}, \\
\frac{\partial e}{\partial t} + {1\over r^2}\frac{\partial}{\partial r}(r^2
e v_r) & = & - \frac{p}{r^2} \frac{\partial}{\partial r}(r^2 v_r)  \nonumber
\\ && + \Gamma_{\rm cr}-\Lambda_{\rm rc}- \Lambda_{\rm gd},
\label{hydro}
\end{eqnarray}
where $\rho$ is the gas density, $v_r$ is the radial velocity, $M$ is the
enclosed mass, $e$ is the internal energy density and $p=e(\gamma_{\rm
i}-1)$ is the gas pressure. We use the ratio of specific heats $\gamma_{\rm
i}=5/3$ to link pressure and internal energy density and modify the
energy  equation for the effects of cooling and heating.  The details
of the radiative cooling rate $\Lambda_{\rm rc}$, gas-dust energy transfer
rate $\Lambda_{\rm gd}$, and cosmic ray heating $\Gamma_{\rm cr}$ are given
in \S~\ref{NIC}. 
The gas dynamics of a collapsing cloud is followed by solving the
usual set of hydrodynamic equations in spherical coordinates 
using the method of finite-differences with a time-explicit, operator split
solution procedure similar to that of the ZEUS-1D numerical hydrodynamics code 
described in detail in Stone \& Norman (\cite{Stone}). 
Because the time scales of cooling and heating are usually much shorter 
than the dynamical time, the energy equation
update due to cooling and heating requires an implicit scheme. Explicit
schemes usually fail due to a strict limitation on the numerical time step
set by the Courant-Friedrich-L{\'e}vy condition.
Therefore, cooling and heating of gas are treated numerically using Newton-Raphson iterations, 
supplemented by a bisection algorithm for occasional zones where the 
Newton-Raphson method does not converge. In order to monitor accuracy, 
the total change in the internal energy density in one time step is kept below 
$15\%$. If this condition is not met, the time step is reduced and a solution 
is again sought.
When the gas number density in the collapsing core 
exceeds $10^{11}$~cm$^{-3}$, we gradually reduce the cooling and heating
so as to establish adiabatic evolution with $\gamma_{\rm i}=5/3$ for $n
>10^{12}$~cm$^{-3}$. 
This simplified treatment of the transition to an opaque protostar
misses the details of the physics on small scales. Specifically,
a proper treatment of the accretion shock and radiative transfer effects is 
required to accurately predict the properties of the stellar
core (see Winkler \& Newman \cite{Winkler} for a detailed treatment and
review of work in this area). 
However, our method should be adequate to study the protostellar
accretion rate, and has been used successfully by e.g.
Foster \& Chevalier (\cite{FC}) and Ogino et al. (\cite{Ogino})
for this purpose.
The numerical grid has 600 points which are initially uniformly spaced,
but then move with the gas until the central
core is formed. This provides an adequate resolution throughout
the simulations.

We impose boundary conditions such that the gravitationally bound cloud core
has a constant mass and constant volume.
The assumption of a constant mass appears to be observationally 
justified. For instance, the observations by Bacmann et al. (\cite{Bacmann}) have shown that
in at least some cases, such as L1544 in Taurus, individual pre-stellar clouds
are characterized by {\it sharp edges} defining typical outer radii
$\sim 0.05-0.5$~pc, implying that these prestellar clouds represent {\it
finite reservoirs of mass} for subsequent star formation. 
Physically, this assumption may be justified if the
core decouples from the rest of a comparatively static, diffuse cloud
due to a shorter dynamical timescale
in the gravitationally contracting central condensation than in the
external region. A specific example of this, due to enhanced magnetic 
support in the outer envelope, is found in the models of ambipolar-diffusion 
induced core formation (see, e.g. Basu \& Mouschovias \cite{Basu95}).
The constant volume condition of a collapsing cloud core is mainly an
assumption of a constant radius of gravitational influence of a subcloud
within a larger parent diffuse cloud.

The radial gas density distribution of a self-gravitating isothermal cloud 
that is in hydrostatic equilibrium can be conveniently approximated by a modified
isothermal sphere, with gas density 
\begin{equation}
\rho={\rho_{\rm c} \over 1+(r/r_{c})^2}
\end{equation}
(Binney \& Tremaine \cite{BT}), where $\rho_{\rm c}$ is the central density and
$r_{\rm c}$ is the radial scale length. 
We choose a value $r_{\rm c}=c_{\rm s}/\sqrt{\pi G \rho_{\rm c}}$, so that the inner profile
is close to that of a Bonnor-Ebert sphere, $\rc$ is comparable
to the Jeans length, and the asymptotic density profile is twice the
equilibrium singular isothermal sphere value $\rho_{\rm SIS} =
\cs^2/(2 \pi G r^2$). The latter is justified on the grounds that 
core formation should occur in a somewhat non-equilibrium manner
(an extreme case is the Larson-Penston flow, in which case the
asymptotic density profile is as high as $4.4 \, \rho_{\rm SIS}$),
and also by observations of protostellar envelope density profiles
that are often overdense compared to $\rho_{\rm SIS}$ (Andr\'e,
Motte, \& Belloche \cite {Andre}). 
We also add a (moderate) positive density perturbation of
a factor $\alpha=1.7$ (i.e. the initial gas density distribution is increased 
by a factor of 1.7) to drive the cloud (especially the inner region 
which is otherwise near-equilibrium) into gravitational collapse. 
For the purpose of comparison with
isothermal simulations, we determine the size of the central flat region
assuming a constant temperature $T=10$~K, which yields $r_{\rm c}=0.03$~pc.

The choice of central density $\rho_{\rm c}$, the size of the central flat
region $r_{\rm c}$, the amplitude of density perturbation $\alpha$, 
and outer radius $r_{\rm out}$ determines the cloud mass and the radial density 
distribution.
The initial gas temperature distribution is then obtained
by solving the equation of thermal balance $\Lambda_{\rm rc}+\Lambda_{\rm
gd}=\Gamma_{\rm cr}$. We study many different cloud masses - six models are 
presented in this paper. Table~\ref{Table1} shows the
parameters for these model clouds. 
The adopted central number density $n_{\rm c}=8.5\times 10^{4}$~cm$^{-3}$
is roughly an order of magnitude lower than is observed in prestellar cores
(Ward-Thompson et al. \cite{WT}). Considering that the observed cores may already be
in the process of slow gravitational contraction, our choice of $n_{\rm c}$
is justified for the purpose of describing the basic features of 
star formation. 
In all six models, the outer radius $r_{\rm out}$ is chosen so as
to form gravitationally unstable prestellar cores with central-to-surface
density ratio $\rho_{\rm c}/\rho_{\rm out}> 14$ (since our initial states
are similar to Bonnor-Ebert spheres). 
In models~NI1, NI3, and NI5, $\rho_{\rm c}/\rho_{\rm out}\approx 19.4$
and by implication $r_{\rm out}/r_{\rm c} \approx 4.3$, whereas in models~NI2,
NI4, and NI6  $\rho_{\rm c}/\rho_{\rm out}\approx 173$ and $r_{\rm out}/r_{\rm c} \approx 13.2$.
Models~NI2, NI4, and NI6 thus represent very extended prestellar cores;
the `NI' stands for nonisothermal.

\begin{table}
\caption{\label{Table1}Model parameters}
\vskip 0.1cm
\begin{tabular}{lllllll}
\hline
\hline
Model & $n_{\rm c}$$^1$ &  $r_{\rm out}$ &${\rho_{\rm c}/
\rho_{\rm out}}$ & $r_{\rm out}/r_{\rm c}$ & $M_{\rm cl}$ & 
$T_{\rm d}$ \\ [2 pt]
\hline
NI1 & $8.5$ & 0.13 & 19.4 & 4.3 & 5 &  -- \\ 
NI2 & $8.5$ & 0.4  & 173  & 13.2 & 20 &  -- \\ 
NI3 & $8.5$ & 0.13 & 19.4 & 4.3 & 5 & 10 \\
NI4 & $8.5$ & 0.4  & 173  & 13.2 & 20 & 10 \\
NI5 & $8.5$ & 0.13 & 19.4 & 4.3 & 5 & 6 \\
NI6 & $8.5$ & 0.4  & 173  & 13.2 & 20 & 6 \\ 
\hline
\end{tabular}
\begin{list}{}{}
\item[$^1$] All densities are in units of $10^4$~cm$^{-3}$, scales in pc, 
masses in $M_\odot$, and temperatures in K. $M_{\rm cl}$ is the cloud mass, 
and $T_{\rm d}$ is the dust temperature.
\end{list}
\end{table}

\section{Results of cloud collapse}
\label{NIC}

In this section we study how the non-isothermality of gas can affect the
temporal evolution of the mass accretion rate. Our results should be interpreted
in the context of models of one-dimensional radial infall. Hence, our calculated
mass accretion rates really represent the infall onto the inner protostellar
disk that would be formed due to rotation. Since disk masses
are not observed to be greater than protostellar masses, it is likely that
the protostellar accretion is at least proportional to the mass infall
on to the disk.
The mass accretion rate is computed at a radial distance 600~AU=0.003~pc.\footnote{We note that the
accretion rate is not expected to vary significantly in the range 0.1~AU -1000~AU, 
according to radiation hydrodynamic simulations of spherical collapse by 
Masunaga \& Inutsuka (\cite{MI}).} We neglect heating due to the central
source at this distance. First, we consider a simplified
assumption of optically-thin gas heated only by the cosmic rays. 
We further take into account the saturation of the radiative cooling of gas 
at $n>10^4$~cm$^{-3}$ and the gas-dust energy transfer (Goldsmith \cite{Gold}).

\subsection{Optically thin limit}
\label{thin}
Our optically thin model serves as an example limiting case that
builds our intuition about the effect of temperature gradients in 
a cloud core and serves as a comparison to the more realistic
cooling models presented in \S\ \ref{thick}.
In this idealized limit, the radiative cooling of gas
can be expressed as 
\begin{equation}
\Lambda_{\rm rc}={\cal L}(T)~n^2  \:\:\: {\rm ergs~cm^{-3}~s^{-1}}, 
\label{cool}
\end{equation}
where ${\cal L}(T)$ is the radiative 
cooling rate in ergs~cm$^3$~s$^{-1}$ and $n$ is the number density.
For the temperature dependence of the radiative cooling rate ${\cal L}(T)$
we have implemented the cooling function of Wada \& Norman (\cite{WN}) (their Fig.~1)  
for solar metallicity and gas temperature $T_{\rm g}<10^4$~K. The 
cooling function simplifies the implementation of cooling by 
collecting the effects of various coolants.  The cooling processes taken into 
account are: 
(1) vibrational and rotational excitation of $H_2$; 
(2) atomic and molecular cooling due to fine-structure emission of C
and O; (3) rotational line emission of CO.
We do not consider cooling due to C$^{+}$, since the complete conversion
of C$^{+}$ to CO in cloud cores occurs when the number density rises beyond
$10^4$~cm$^{-3}$ (Nelson \& Langer \cite{NL}). Moreover, this chemical
change appears to have little effect on the dynamics of self-gravitating
cloud cores. This is because the cooling rate of C$^{+}$ is similar to that
of CO, particularly in the density-temperature region typical for the cloud
cores (Nelson \& Langer \cite{NL}).

Since we are interested in well-shielded regions, we can ignore the direct 
photoelectric heating of gas by the incident ultraviolet radiation.
The cosmic-ray heating is thus the only process that heats the gas directly.
We adopt the cosmic-ray heating per unit volume (Goldsmith \cite{Gold})
\begin{equation}
\Gamma_{\rm cr}=10^{-27} \: \left( {n \over {\rm cm^{-3}}} \right)  \:\:\:  {\rm ergs \: cm^{-3}\: s^{-1} }.
\label{heat}
\end{equation}

\begin{figure}
  \resizebox{\hsize}{!}{\includegraphics{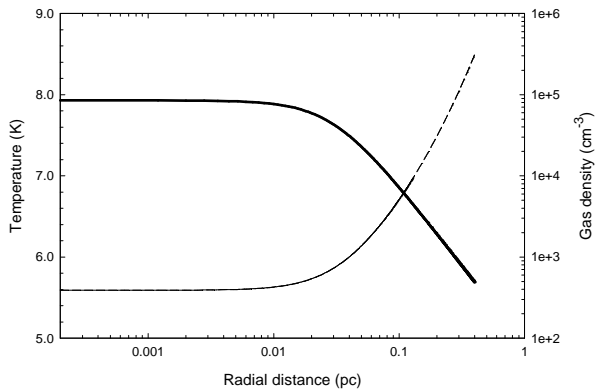}}
      \caption{The initial radial gas temperature distribution
      for the optically thin models NI1 (solid line) and NI2 (dashed line)
      listed in Table~\ref{Table1}. Note that the temperatures for
      the two models are identical for $r<0.13$~pc. The thick solid line 
      plots the initial radial distribution of the gas density.
      }
         \label{fig1}
\end{figure}

The initial gas temperature distribution is obtained
by solving the equation of thermal balance $\Lambda_{\rm rc}=\Gamma_{\rm
cr}$ for the adopted radial gas density profile of model~NI1. The solid
line in Fig.~1 shows the resulting gas temperature $T_{\rm g}$. 
It varies from 5.6~K in the cloud's center 
to 7~K at the outer edge for the cloud with $r_{\rm out}=0.13$~pc. For a larger cloud 
of radius $r_{\rm out}=0.4$~pc (model NI2),
the temperature grows from 5.6~K in the center to 8.5~K at the outer edge,
as shown by the dashed line in Fig.~\ref{fig1}.

\begin{figure}
  \resizebox{\hsize}{!}{\includegraphics{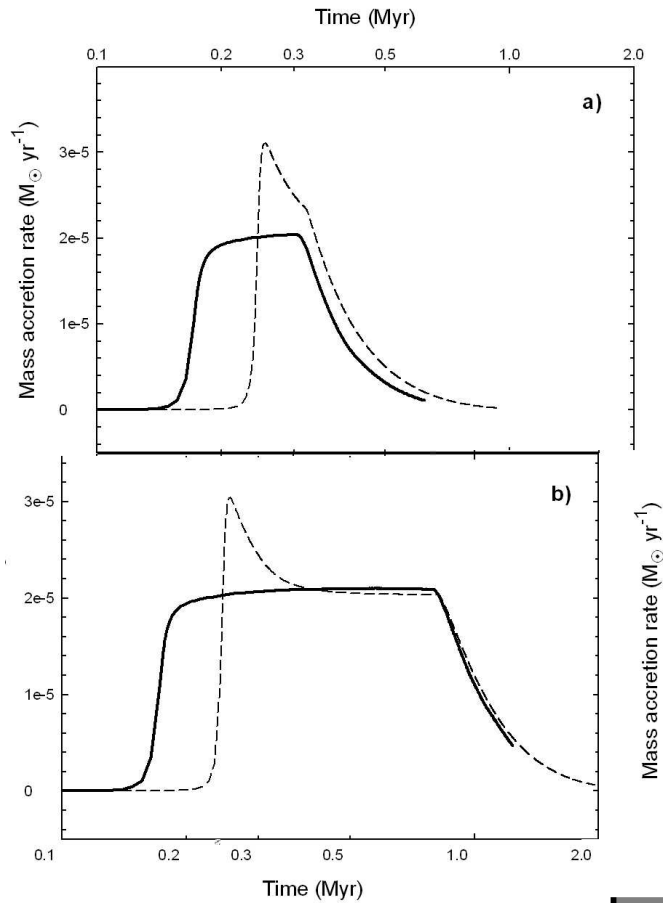}}
      \caption{
      The temporal evolution of the mass accretion rates (solid lines) obtained in
      {\bf a)} model NI1 and {\bf b)} model~NI2. 
      The dashed lines give the mass accretion rates for the corresponding
      isothermal ($T_{\rm g}=10~K$) clouds.}
         \label{fig2}
\end{figure}

After the collapse is initiated, some aspects of the temporal evolution
are shown in Fig.~2.
The solid lines in Figs.~\ref{fig2}a and \ref{fig2}b show 
the evolution of the mass accretion rate  $\dot{M}$  at a radial distance of 600~AU from the center
for models~NI1 and NI2, respectively. The accretion
rates of the corresponding isothermal clouds ($T_{\rm g}=10$~K) are plotted 
by the dashed lines for comparison.
An obvious difference is seen in the behaviour of $\dot{M}$
of the non-isothermal clouds
as compared to that of the isothermal ones: {\it no peak in the mass 
accretion rate is seen at the time of the hydrostatic core formation} ($t \approx 0.18$~Myr).
This is because the effect of progressively warmer layers falling onto the 
protostar cancels the effect of inner mass shells having greater initial
infall speeds; we discuss this effect again in \S\ \ref{mdot}. 
Instead, $\dot{M}$ of the non-isothermal clouds grows monotonically 
and appears to stabilize at $\sim 2.1\times 10^{-5}~M_\odot$~yr$^{-1}$
if the size of the cloud is sufficiently large (model~NI2).
Note that the positive temperature gradient developing in 
the non-isothermal pre-collapse 
clouds shortens the duration of the pre-core-formation phase by  $\sim 0.07$~Myr
as compared to that of the isothermal clouds.
A sharp drop of the mass accretion rate at $\approx 0.31$~Myr in model~NI1 
(at $\approx 0.83$~Myr in model~NI2) is due to a rarefaction wave propagating inward from the
cloud's outer edge (see Vorobyov \& Basu \cite{VB} for details). 
In case of a larger cloud of $r_{\rm out}=0.4$~pc (model~NI2), it takes
a longer time of $\sim 0.8$~Myr for the rarefaction wave to reach the innermost
regions. As a consequence, model~NI2 exhibits a longer period of nearly-time-independent
mass accretion rate than model~NI1. At the time when the rarefaction wave
reaches the radius of $r \sim 600$~AU, roughly half of the envelope mass
has accreted on to the central hydrostatic core.

\subsection{Cooling saturation at higher densities}
\label{thick}
Calculations of thermal balance in dark, well-shielded molecular
cores by Goldsmith \& Langer (\cite{Gold2}) and 
Goldsmith (\cite{Gold}) have shown that the cooling from main coolants
such as CO, $\rm C^{\rm 13}O$, $\rm H_{\rm 2}O$, C, and others, 
is proportional to $n^2$ only at low densities
of molecular hydrogen, i.e. $n \le 10^3-10^4$~cm$^{-3}$.
At $n \ge 10^4$~cm$^{-3}$, 
the cooling from these species saturate (see e.g. Goldsmith \cite{Gold}). 
This is because the transitions
that can contribute to the cooling are thermalized due to the combination
of a higher collision rate and radiative trapping. Noticeable exceptions
are the cooling from $\rm C^{\rm 18}O$ and CS, which do not saturate even
at densities $n \sim 10^7$~cm$^{-3}$ due to a sufficiently
large spontaneous decay rate.
As a result, the total cooling has a complicated dependence on the temperature
and density of molecular hydrogen, expressed by Goldsmith (\cite{Gold}) as
\begin{equation}
\Lambda_{\rm rc}=\alpha \left({T_{\rm g} \over 10~{\rm K}} \right)^{\beta}
\:\:\: {\rm ergs~cm^{-3}~s^{-1}},
\end{equation}
where $\alpha$ and $\beta$ are given in Table~2 of Goldsmith (\cite{Gold})
for the undepleted (standard) molecular abundances.

As in \S~\ref{thin}, we assume that the cosmic-ray heating is the only
mechanism that heats the gas directly.
The dust grains heated by the diffuse visual-IR radiation field may be
an additional source of indirect gas heating/cooling depending on the difference between
the gas and dust temperatures. For the gas-dust energy transfer, we adopt
the expression given in Goldsmith (\cite{Gold}),
\begin{eqnarray}
\Lambda_{\rm gd}&=&2\times 10^{-33} \left( {n \over {\rm
cm^{-3}}}  \right)^2 \left(  {T_{\rm g} - T_{\rm d} \over {\rm K}}  \right)
 \nonumber \\
& & \times \left( {T_{\rm g} \over 10~{\rm K}}  \right) \:\:\: {\rm ergs\: cm^{-3}\: s^{-1}},
\end{eqnarray}
where $T_{\rm g}$ and $T_{\rm d}$ are the temperatures of gas and dust,
respectively. When the dust temperature is greater than that of the gas,
the dust heats the gas and vice versa. 

\subsubsection{Initial gas temperature profile}
\label{tempprofile}
The radial distribution
of the gas temperature depends on the adopted temperature of dust.
As demonstrated by Zucconi et al. (\cite{Zucconi}), 
the dust temperature in the prestellar cores is sensitive to the 
center-to-edge visual extinction $A_{\rm v}$ of a parent cloud. 
We consider a simple model in which the dust temperature is
constant throughout the pre-stellar core and equal to 6~K or 10~K. 
These two limits correspond to cores that are heavily shielded ($A_{\rm v}\sim
100^{m}.0$) or only moderately shielded ($A_{\rm v} \sim 5^{m}.0$) from the external radiation. 
\begin{figure}
  \resizebox{\hsize}{!}{\includegraphics{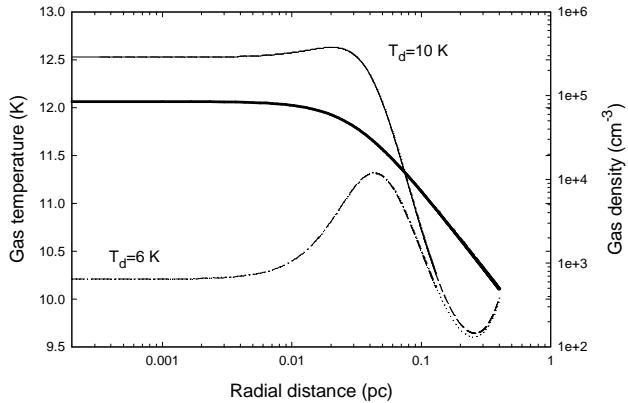}}
      \caption{The initial radial gas temperature distributions 
      in four models listed in Table~\ref{Table1}: the thin solid line
      -- model~NI3, the dashed line -- model~NI4; the dotted-dashed line
      -- NI5, and the dotted line -- model~NI6. Note that the radial distributions
      of the gas temperature for models~NI3 and NI4 and models~NI5 and
      NI6 merge at $r<0.13$~pc. The thick solid line plots the initial radial
      distribution of the gas density. }
         \label{fig3}
\end{figure}

The initial temperature distribution of gas is obtained
by solving the equation of thermal balance $\Lambda_{\rm rc}
+\Lambda_{\rm gd}=\Gamma_{\rm cr}$. In model~NI3,
the gas temperature $T_{\rm g}$ decreases from 12.6~K  
near the cloud's center to 10.3~K at the outer edge ($r_{\rm
out}=0.13$~pc) as shown in Fig.~\ref{fig3} by the thin solid line. 
The development of the negative temperature gradient in the 0.03-0.13~pc range is due to 
radiative cooling saturation at higher densities $n>10^4$~cm$^{-3}$, where
the cosmic ray heating dominates the radiative cooling.
At the same time, the gas-dust energy transfer
becomes an efficient coolant at $n>10^5$~cm$^{-3}$ and the gas
temperature stabilizes at $T_{\rm g}\approx 12.6$~K in the innermost region
$r<0.02$~pc.
In the case of a larger cloud with $r_{\rm out}=0.4$~pc (model~NI4), the gas temperature
exhibits a more complicated behaviour as shown in Fig.~\ref{fig3} by the
dashed line (note that it merges with the thin solid line at $r<0.13$~pc). 
It decreases from 12.6~K
near the cloud's center and  reaches a minimum value of 9.6~K at $r\sim 0.25$~pc.
Further out, the gas temperature grows again to 10.0~K at $r_{\rm out}=0.4$~pc.
The latter indicates that the gas becomes effectively optically thin at
$n \le 10^3$~cm$^{-3}$ and the radial temperature distribution
of gas develops a positive temperature gradient, as was indeed found in \S~\ref{thin}
for the optically thin radiative cooling.

For a lower value of the dust temperature $T_{\rm d}=6$~K, the 
initial radial distribution of gas temperature is shown in Fig.~\ref{fig3} by the dotted-dashed and
dotted lines for the clouds of $r_{\rm out}=0.13$~pc (model~NI5)
and $r_{\rm out}=0.4$~pc (model~NI6), respectively.
The dust-gas energy transfer reduces the gas temperature in the
cloud's innermost regions by $\approx 2$~K as compared to the previously considered
case of $T_{\rm d}=10$~K. At the lower densities, the gas temperature
is mainly determined by the balance of radiative cooling of gas and 
cosmic ray heating. As a consequence, the gas temperature profiles 
for both adopted values of $T_{\rm d}$ become very similar for
$n<10^4$~cm$^{-3}$.

\subsubsection{Mass accretion rate}
\label{mdot}

\begin{figure}
  \resizebox{\hsize}{!}{\includegraphics{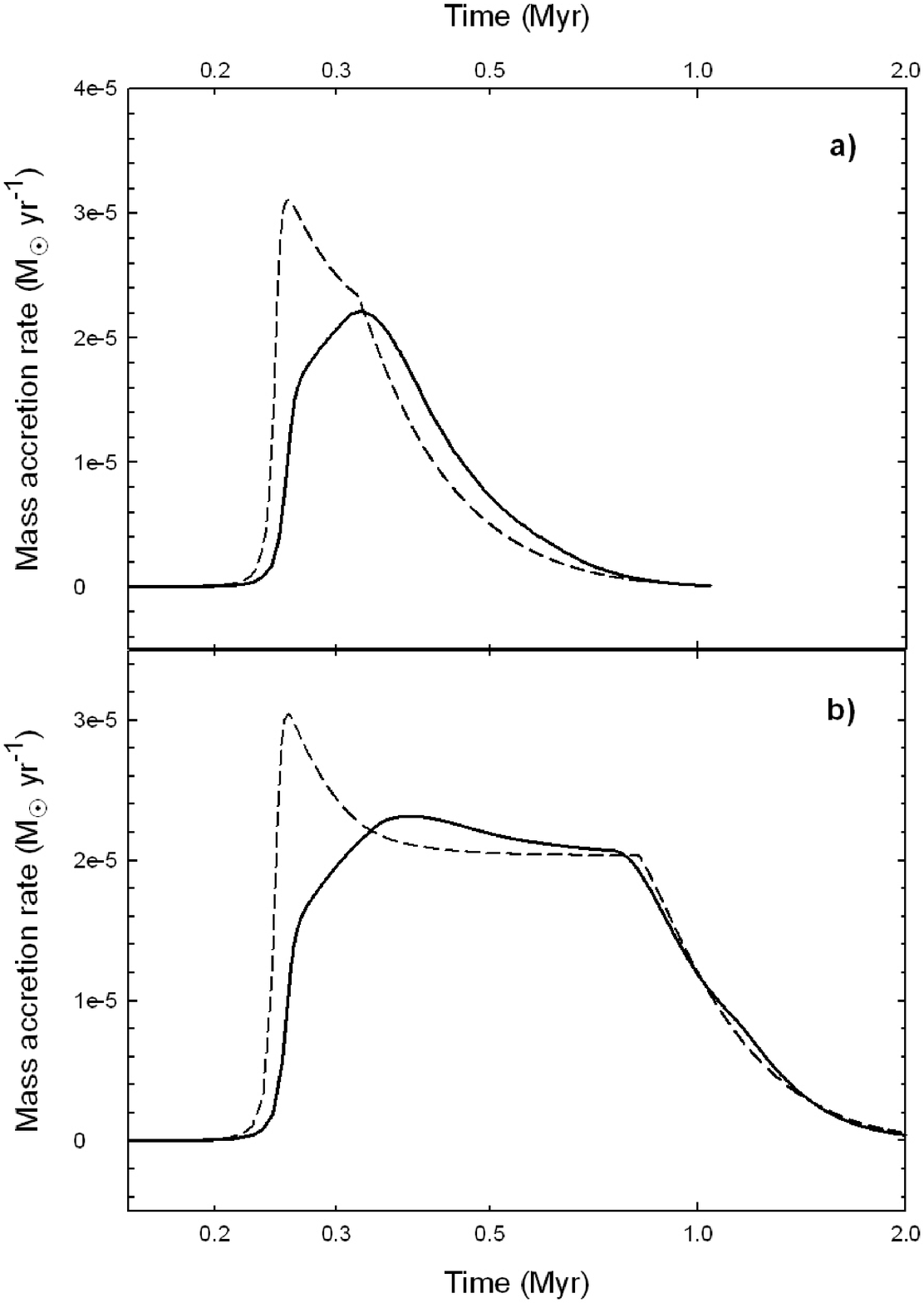}}
      \caption{The temporal evolution of the mass accretion rate 
      shown with the thick solid line for {\bf a)} models~NI5 and 
      {\bf b)} NI6, respectively.
      Both models have a dust temperature $T_{\rm d} = 6$ K.
      The dashed lines show the accretion rate for the corresponding
      isothermal ($T_{\rm g}=10$~K) clouds.}
       \label{fig4}
\end{figure}

\begin{figure}
  \resizebox{\hsize}{!}{\includegraphics{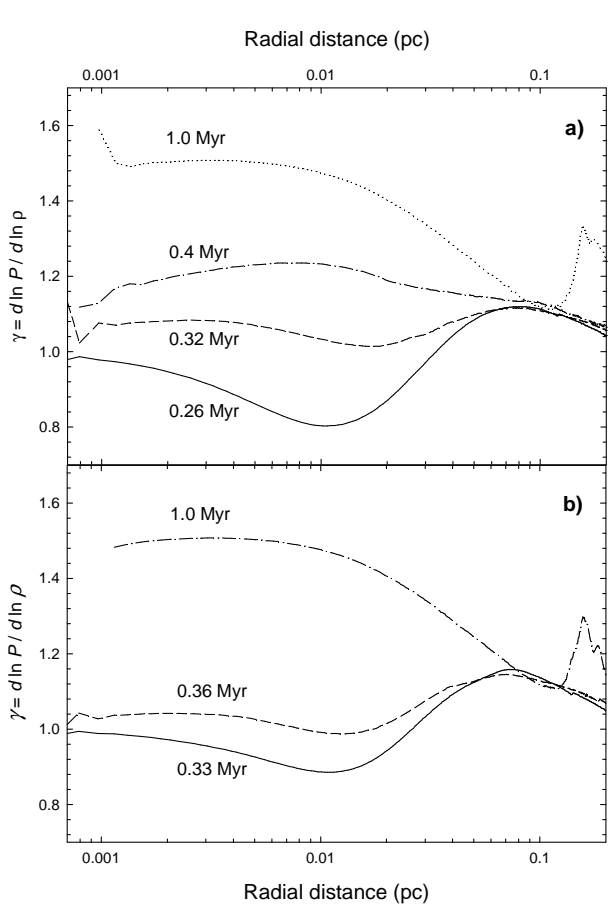}}
      \caption{Radial profiles of the local polytropic index $\gamma$ obtained
      for {\bf a)} model~NI6 ($T_{\rm d}=6$~K) and {\bf b)} model~NI4 ($T_{\rm
      d}=10$~K) at different evolutionary times as indicated in each panel.}
       \label{fig5}
\end{figure}

\begin{figure}
  \resizebox{\hsize}{!}{\includegraphics{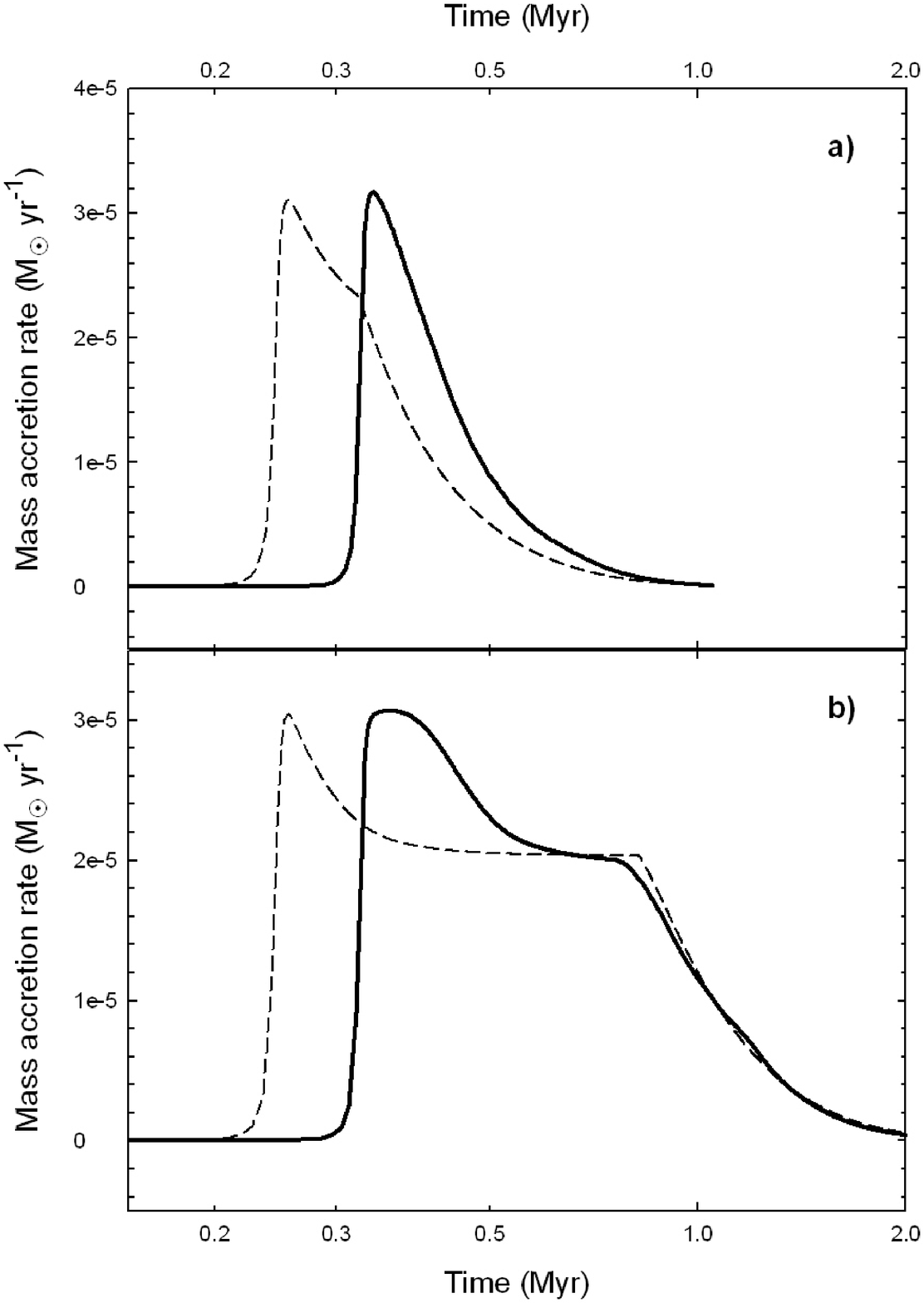}}
      \caption{The temporal evolution of the mass accretion rate 
      shown with the thick solid line for {\bf a)} models~NI3 and 
      {\bf b)} NI4, respectively.
      Both models have a dust temperature $T_{\rm d} = 10$ K.
      The dashed lines show the accretion rate for the corresponding
      isothermal ($T_{\rm g}=10$~K) clouds. }
       \label{fig6}
\end{figure}

The solid lines in Fig.~\ref{fig4}a and \ref{fig4}b 
show the temporal evolution of the mass accretion rate $\dot{M}$ 
for models~NI5 and NI6, respectively.  
The corresponding accretion rates of isothermal ($T_{\rm g}=10$~K) clouds are plotted 
by the dashed lines for comparison. 
{\it At the lower dust temperature $T_{\rm d}=6$~{\rm K}, the accretion rate resembles
that of the optically thin cloud}: $\dot{M}$ 
has no well-developed peak at the time of the hydrostatic 
core formation at $t=0.26$~Myr. Thus, the gas-dust energy exchange acts
as an effective thermostat; the gas collisionally transfers its energy to the dust
instead of directly radiating it by photon emission. We note here that
such low dust temperatures of $T_{\rm d}\sim 6$~K may indeed be present in dense pre-stellar
cores as suggested by the self-consistent modeling of Zucconi et al. (\cite{Zucconi}) and
Galli et al. (\cite{Galli}).

The self-similar solutions for the accretion phase of isothermal cloud 
collapse (Shu \cite{Shu}; Hunter \cite{Hunter2}) predict that the mass
accretion rate is time-independent and that
\begin{equation}
\dot{M} =  k \, {c_{\rm s}^3 \over G},
\label{sim}
\end{equation}
where $c_{\rm s}$ is the isothermal sound speed. The coefficient $k$ is equal 
to 0.975 in the Shu solution,
whereas in the Hunter (\cite{Hunter2}) extension (to the accretion phase) 
of the Larson (\cite{Larson})-Penston (\cite{Penston}) solution, $k = 46.9$. 
This difference is due to the different velocity $v(r)$ and density $\rho(r)$ 
radial profiles of gas at the time when the central
hydrostatic core forms: they are $v(r)=0$ and $\rho(r)=c_{\rm
s}^2/2\pi G r^2$ in the Shu solution and $v(r)=-3.3~c_{\rm s}$ and 
$\rho(r)=4.4~c_{\rm s}^2/2\pi G r^2$ 
in the Larson-Penston-Hunter solution. However, numerical simulations of 
both isothermal and non-isothermal collapse 
show that the gas velocity at the time of stellar core formation is not constant
with radius; the absolute value of gas velocity is only approaching 
$3.3~c_{\rm s}$ near the hydrostatic core, while decreasing at larger radii
and converging to zero at the outer boundary. 
As a result, the peak in $\dot{M}$ appears right after the formation
of a central hydrostatic stellar core.

\begin{figure}
  \resizebox{\hsize}{!}{\includegraphics{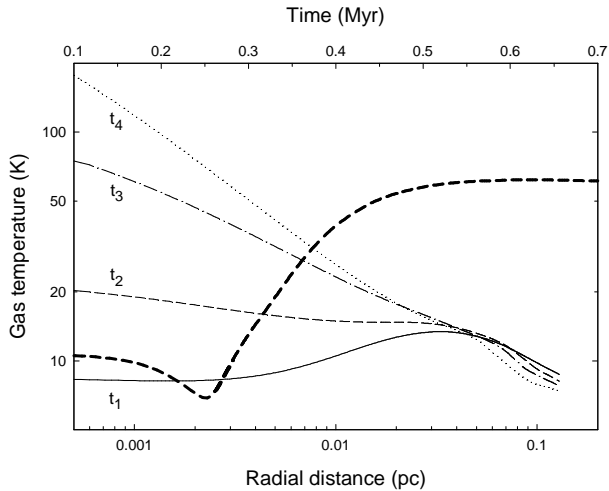}}
      \caption{The radial gas temperature distribution obtained in model~NI5
      at four different evolutionary times: $t_1=0.27$~Myr -- the solid line,
      $t_2=0.32$~Myr -- the thin dashed line, $t_3=0.4$~Myr -- the dotted-dashed
      line, and $t_4=0.5$~Myr -- the dotted line. The thick dashed lines
      plots the temporal evolution of the gas temperature at $r=600$~AU
      from the center of the cloud.}
         \label{fig7}
\end{figure}

The absence of the peak in models~NI5 and NI6
can be understood if one considers the {\it local}
polytropic index $\gamma = d \ln P/ d \ln \rho$ obtained from the model's
known radial distributions of pressure and density.  Fig.~\ref{fig5}a plots
the radial distribution of $\gamma$ obtained for model~NI6 
at four different evolutionary times.  
Shortly after the formation of the hydrostatic core at $t \approx 0.26$~Myr, the
gas flow in the inner 0.04~pc is characterized by $0.8\la\gamma <1.0$. 
Since the polytropic law implies that
$T_{\rm g}~\rho^{1-\gamma}={\it constant}$ during the collapse, and the
gas density in the infalling envelope decreases with radius as $r^{-1.5}$, 
the gas temperature $T_{\rm g}$ and the associated sound speed $c_{\rm s}$ 
will increase with radius as long as $\gamma<1$.
Because the similarity solutions (see Eq.~\ref{sim}) predict that $\dot{M}$ 
is directly proportional to the sound speed,
the mass accretion rate will increase with time, as progressively warmer
layers of gas (characterized by increasing $c_{\rm s}$)  
fall onto the central hydrostatic core. 
As a result, {\it the increase of $\dot{M}$ due to the strong positive temperature gradient 
appears to compensate the decrease in $\dot{M}$ due to the non-constant
radial velocity profile} and the mass accretion rate in this early phase
becomes a monotonically growing function of time as shown by the 
solid lines in Fig.~\ref{fig4}a.
At $t\approx0.32$~Myr the polytropic index in the infalling envelope becomes
greater than unity throughout the infalling envelope.
At roughly the same time, the accretion rate attains the maximum value of $2.35\times
10^{-5}~M_\odot$~yr$^{-1}$. Further evolution is characterized by a slowly
declining $\dot{M}$. In this
intermediate phase, the mass accretion rate appears to converge to that of
the isothermal cloud. 
The late phase of accretion ($t>0.8$~Myr) is characterized by a rapid and terminal
decline when the rarefaction wave caused by a finite mass reservoir arrives
at the center. This effect was discussed in detail in paper~I.
The transition from the early phase to late phase accretion occurs smoothly
for the smaller clouds (i.e. model~NI5), but the intermediate phase of
slowly declining mass accretion is still clearly
visible for larger clouds in which the effect of the rarefaction wave is  
delayed due to a larger radius.

The solid lines in Fig.~\ref{fig6}a and \ref{fig6}b 
show the temporal evolution of the mass accretion rate $\dot{M}$ 
for models~NI3 and NI4, respectively.  
The corresponding accretion rates for isothermal ($T_{\rm g}=10$~K) clouds are plotted 
by the dashed lines. At a higher dust temperature $T_{\rm
d}=10$~K, the accretion rate resembles that of the isothermal cloud: a
well-developed peak is visible at the time of hydrostatic core formation $t\approx0.33$~Myr. 
Again, as in the case of lower dust temperature, we consider the radial
profiles of local polytropic index obtained from the model's known pressure and density
profiles. Fig.~\ref{fig5}b shows the radial distribution of $\gamma$ 
in model~NI4 at three times. Shortly after
the formation of hydrostatic core at $t\approx 0.33$~Myr,
the gas flow in the inner envelope $r<0.03$~pc has local polytropic index
that is only slightly below unity ($0.9 \la \gamma \la 1.0$).
As a consequence, {\it the very weak
positive temperature gradient is not capable of compensating the
decrease in the mass accretion rate due to the non-constant radial velocity profile}
and the behaviour of the accretion rate
is qualitatively similar to that of the isothermal cloud.
At $t\approx 0.36$~Myr the polytropic
index grows above unity in the infalling envelope, as is seen from the dashed 
line in Fig.~\ref{fig5}b. In this phase, the accretion rate declines due
to both the combined action of cooling and heating and the non-constant radial
velocity profile developed in the prestellar phase.
However, as in the case of lower dust temperature,
a sharp and terminal decline of accretion at the late protostellar stages
$t>0.8$~Myr is due to the rarefaction wave rather than to any other effect.

\subsubsection{Evolution of gas temperature}
\label{tempevol}

In the post-core-formation
epoch,  the gas is virtually in a free-fall motion near the central hydrostatic
core and the characteristic dynamical time becomes comparable or even shorter
than that of the cooling. 
As a result, the flow becomes nearly adiabatic with $\gamma \sim 1.5$ and the gas 
temperature in the central region grows with time. The latter tendency is clearly seen in
Fig.~\ref{fig7}, where we plot the radial gas temperature distribution
in model~NI5 obtained at four different evolutionary times.  
The initial radial gas temperature distribution
shown by the dotted-dashed line
in Fig.~\ref{fig3} is characterized by a mild increase towards the
outer edge of a cloud, so that it reaches a maximum of $T_{\rm g}\sim 11.3$~K 
at $r \sim 0.05$~pc
and decreases to roughly the central value of $T_{\rm g}=10.2$~K at the outer
edge $r=0.13$~pc. This behaviour of the gas temperature remains qualitatively
similar until and
shortly after the formation of the central hydrostatic core at $t\approx 0.26$ Myr.
However, in the post-core-formation stage, the radial
distribution of the gas temperature develops a negative gradient, the
slope of which gradually grows with time as shown in Fig.~\ref{fig7} by  
the thin dashed, 
dotted-dashed and dotted lines  for $t=0.32$, 0.4, and
0.5~Myr, respectively. The development of the negative temperature
gradient is due to compressional heating $P (\bl \nabla \cdot  \bl v)$
that overtakes other heating/cooling processes at $r<0.04$~pc.
We note that due to some residual cooling in the interior of the infalling
envelope, the polytropic index $\gamma$ tends to a value of 3/2 instead
of 5/3.
The thick dashed line in Fig.~\ref{fig7} shows the temporal evolution of 
the gas temperature at the radial distance of 600~AU from the cloud's center. 
The initial decrease of the gas temperature prior to the formation of the central
hydrostatic core is followed by an increase and then a saturation at $T_{\rm
g} \sim 60$~K in the post-core-formation stage.

\section{Conclusions}
\label{Sum}
We have numerically followed the gravitational collapse of spherical pre-stellar
cores of finite mass and volume down to the protostellar stage. 
The influence of the cooling and heating on the temporal evolution of 
the mass accretion rate $\dot{M}$ has been investigated.
We summarize our results as follows.

(i) {\it Optically thin gas heated only by cosmic
rays}. In a simplified approach where the radiative cooling of gas
is proportional to the square of the gas density, the mass accretion rate monotonically
increases with time and appears to attain a constant value  
after the formation of the hydrostatic
core, if the size of a cloud is sufficiently
large. This temporal evolution of $\dot{M}$ is qualitatively different
from that found in isothermal simulations, which show the development
of a peak and subsequent decline in $\dot{M}$ shortly after the 
formation of the hydrostatic core 
(see e.g. Hunter \cite{Hunter2}; Foster \& Chevalier \cite{FC}; 
Ogino et al. \cite{Ogino}; paper I).
The later evolution of the mass accretion rate, after roughly half of the
envelope has accreted on the central hydrostatic core,
shows a sharp drop due to the gas rarefaction wave propagating inward from the cloud's outer edge.
This drop in $\dot{M}$ is a direct consequence of the assumed finite mass 
and volume of the gravitationally bounded non-isothermal cloud.

(ii) {\it The effects of radiative 
cooling saturation and gas-dust energy transfer}. 
In a more realistic approach where the radiative cooling of 
the gas saturates
at  $n \ge10^{4}$~cm$^{-3}$ (Goldsmith \cite{Gold}) and the gas-dust
energy transfer is taken into account,
the temporal evolution of the mass accretion rate depends on the dust temperature.
If the dust temperature is sufficiently low ($T_{\rm d} \sim 6$~K), which
roughly corresponds to the gravitationally bounded subcloud 
being deeply embedded within a parent diffuse (i.e. non-gravitating) 
molecular cloud  and heavily shielded from the interstellar radiation field
($A_{\rm v} \sim 100^m.0$), the gas behaves as being effectively optically thin
due to efficient cooling by dust. Specifically, no well-developed
peak in $\dot{M}$ is observed shortly after the formation of the hydrostatic
core, and the temporal evolution of the mass accretion rate is qualitatively
similar to that considered in the optically thin case.
The absence of the peak is due to the positive temperature
gradient developing in the cloud's innermost regions 
before and shortly after the formation of the hydrostatic core.
This positive temperature gradient effectively increases the mass accretion 
rate with time and appears to compensate its decrease due to the non-uniform radial
gas velocity profile.
In the opposite case of a subcloud being only moderately shielded 
from the interstellar radiation field with $A_{\rm v} \sim 5^m.0$ and 
$T_{\rm d} \sim 10$~K, the temporal evolution of the mass accretion rate
is qualitatively similar to that of the isothermal ($T_{\rm g}=10$~K) cloud.

Irrespective of the effects of radiative cooling and dust temperature,
the later evolution of the mass accretion
rate, after roughly half of the envelope has accreted onto the hydrostatic
core, is characterized by a fast decline due to a rarefaction wave associated
with the finite mass reservoir.
In paper I, we associated this phase (which has a temporally declining
bolometric luminosity $L_{\rm bol}$) with the empirically defined
class I phase of protostellar evolution. 
The inclusion of non-isothermal effects in this paper does not change
this conclusion. We have shown that the initial decline in 
$\dot{M}$ is weakened or may even be absent
in a model with a realistic treatment of cooling.
Therefore, the ultimate decline in $\dot{M}$ due to the finite mass
reservoir is even more necessary to explain the class I phase.

Our simulations also demonstrate 
that the evolution of pre-stellar cores down to the late protostellar
stage cannot be described by a polytropic law with a 
fixed index $\gamma = d \ln P / d\ln \rho$. Instead, before 
and shortly after the formation
of the hydrostatic core, the local polytropic index $\gamma$ in the inner
envelope $r<0.03$~pc is below unity and 
$\sim 0.8-1.0$, mainly due to effective cooling
by the dust. However, in the post-core-formation epoch, the gas is virtually
in a free-fall near the central hydrostatic core and the compressional 
heating overtakes other energetic processes. As a consequence,
the flow becomes nearly adiabatic with $\gamma \sim 1.5$ and the gas in the accreting
envelope attains a negative temperature gradient.

\section*{Acknowledgments}
EIV gratefully acknowledges present support
from a CITA National Fellowship and past support by the NATO Science Fellowship
Program administered by the Natural Sciences and Engineering Research Council
(NSERC) of Canada. SB was supported by a research grant from NSERC.

\end{document}